\documentclass[preprint]{revtex4}

\usepackage{graphicx}
\usepackage{epsfig} 
\usepackage{dcolumn}
\usepackage{bm}
\usepackage{subfigure} 

\newcommand{\eqref}[1]{(\ref{#1})}
\newcommand{\order}[1]{\mathcal{O}\left(#1\right)}

\newcommand{\avg}[1]{\left\langle#1\right\rangle}
\newcommand{\paren}[1]{\left(#1\right)}
\newcommand{\bracket}[1]{\left\lbrack#1\right\rbrack}
\newcommand{\abs}[1]{\left|#1\right|}

\newcommand{\br}{\bm{r}}
\newcommand{\re}[1]{\textrm{Re}\left(#1\right)}
\newcommand{\im}[1]{\textrm{Im}\left(#1\right)}

\begin{document}

\title{Chimera States in a Ring of Nonlocally Coupled Oscillators}

\author{Daniel M. Abrams and Steven H. Strogatz}
\affiliation{Niels Bohr Institute, Blegdamsvej 17, 2100 Copenhagen
\O, Denmark\footnote{Permanent Address: 212 Kimball Hall,
Department of Theoretical and Applied Mechanics, Cornell \\
University, Ithaca, NY 14853-1503, USA}}

\begin{abstract} Arrays of identical limit-cycle oscillators have
been used to model a wide variety of pattern-forming systems, such
as neural networks, convecting fluids, laser arrays, and coupled
biochemical oscillators.  These systems are known to exhibit rich
collective behavior, from synchrony and traveling waves to
spatiotemporal chaos and incoherence.  Recently, Kuramoto and his
colleagues reported a strange new mode of organization---here
called the chimera state---in which coherence and incoherence
exist side by side in the same system of oscillators.  Such states
have never been seen in systems with either local or global
coupling; they are apparently peculiar to the intermediate case of
nonlocal coupling.  Here we give an exact solution for the chimera
state, for a one-dimensional ring of phase oscillators coupled
nonlocally by a cosine kernel.  The analysis reveals that the
chimera is born in a continuous bifurcation from a spatially
modulated drift state, and dies in a saddle-node collision with an
unstable version of itself.
\end{abstract}

\maketitle

\section{Introduction}

\subsection{The chimera state}

A fascinating spatiotemporal pattern was reported recently by
Kuramoto, Battogtokh and Shima \cite{kur02, kur03, shima04}. While
studying arrays of identical limit-cycle oscillators that are
coupled nonlocally, they found that for certain choices of
parameters and initial conditions, the array would split into two
domains: one composed of coherent, phase-locked oscillators,
coexisting with another composed of incoherent, drifting
oscillators.  The coexistence of locking and drift was robust.
It occurred in both one and two spatial dimensions, and for
various kinds of oscillators, including the Fitzhugh-Nagumo model,
complex Ginzburg-Landau equations, phase oscillators, and an
idealized model of biochemical oscillators.

It's important to appreciate how unexpected this coexistence state
was. Nothing like it had ever been seen before, at least not in an
array of identical oscillators.  Normally, identical oscillators
settle into one of a few basic patterns
\cite{winfree80,kur84,cross93}. The simplest is synchrony, with
all oscillators moving in unison, executing identical motions at
all times.  Another common pattern is wave propagation, typically
in the form of solitary waves in one dimension, spiral waves in
two dimensions, and scroll waves in three dimensions. The common
feature in these cases is that all the oscillators are locked in
frequency, with a fixed phase difference between them. At the
opposite end of the spectrum is incoherence, where the phases of
all the oscillators drift quasiperiodically with respect to each
other, and the system shows no spatial structure whatsoever.  And
finally, one sometimes sees more complex patterns, including
modulated structures, spatiotemporal chaos, intermittency and so
on.

What was so odd about the coexistence state is that two of these
patterns (locking and incoherence) were present in the same
system, simultaneously. This combination of states couldn't be
ascribed to the simplest mechanism of pattern formation---a
supercritical instability of the spatially uniform
oscillation---because it can occur even if the uniform state is
linearly stable, as indeed it was for the parameter values used by
Kuramoto and his colleagues. Furthermore, it has nothing to do
with the classic partially locked/partially incoherent states that
occur in populations of non-identical oscillators with distributed
natural frequencies \cite{winfree67,kur84}. There, the splitting
of the population stems from the inhomogeneity of the oscillators
themselves; the desynchronized oscillators are the intrinsically
fastest and slowest ones in the tails of the distribution.  In
contrast, for the system studied by Kuramoto et al., there is no
distribution of frequencies.  All the oscillators are the same,
and yet they still break apart into two groups of utterly
different character.

Because the coexistence state involves two seemingly incompatible
forms of dynamical behavior,  we will henceforth refer to it as
``the chimera state," inspired by the mythological creature
composed of a lion's head, a goat's body, and a serpent's tail.
Today the word chimera is used more generally to indicate
something made up of incongruous parts, or something that seems
wildly improbable or fantastical.

\begin{figure}[t] 
    \centering{    \includegraphics[height=5cm]{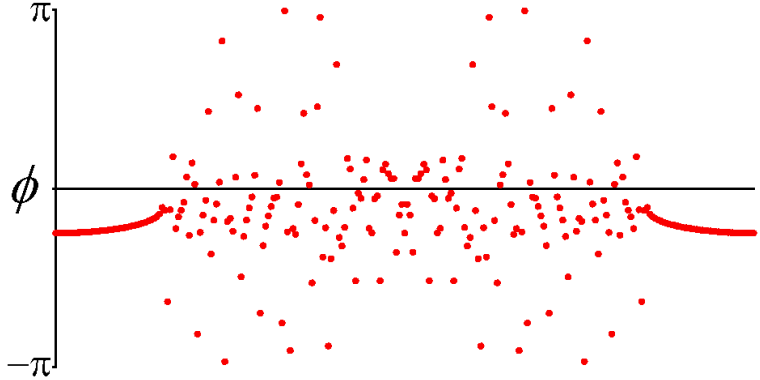}    }
    \caption{Phase pattern for a typical chimera state.  Here $\kappa=4.0$,
        $\alpha=1.45$, $N=256$ oscillators.  Equation \eqref{eq:kurphase} was
        integrated with fixed time step $dt=0.025$ for 8,000 iterations,
        starting from $\phi(x)=6\exp\bracket{-30(x-\frac{1}{2})^2}r(x)$, where $r$ is a uniform
        random variable on $\bracket{-\frac{1}{2}, \frac{1}{2}}$. }
    \label{fig:phiplot}
\end{figure}

Figure \ref{fig:phiplot} shows a realization of the chimera state
in the simplest setting, a one-dimensional ring of phase
oscillators \cite{kur02, kur03}.  The governing equation is
\begin{equation} \label{eq:kurphase}
   \frac{\partial \phi}{\partial t} = \omega -
   \int_{0}^{1}{G\paren{x-x'}\sin \bracket{\phi(x,t) - \phi(x',t) + \alpha} dx'}
\end{equation}
where $\phi(x,t)$ is the phase of the oscillator at position $x$
at time $t$. The space variable $x$ runs from 0 to 1 with periodic
boundary conditions, and should be regarded as an angle on a
circle (mod 1).  The frequency $\omega$ plays no role in the
dynamics, in the sense that one can set $\omega = 0$ without loss
of generality by redefining $\phi \to \phi + \omega t$, without
otherwise changing the form of equation \eqref{eq:kurphase}.  The
kernel $G(x-x')$ provides nonlocal coupling between the
oscillators.  It is assumed to be even, non-negative, decreasing
with the separation $|x-x'|$ along the ring, and normalized to
have unit integral. Specifically,  Kuramoto and Battogtokh
\cite{kur02, kur03} used an exponential kernel $G(x-x') \propto
\exp \paren{-\kappa |x-x'|}$. Then, for parameter values
$\alpha=1.457$ and $\kappa=4$ and suitable initial conditions (to
be discussed in detail in Section \ref{sec:simulation}), the
system evolves to the chimera state shown in
Fig.~\ref{fig:phiplot}.

\begin{figure}[t] 
    \centering{   \includegraphics[height=5.9cm]{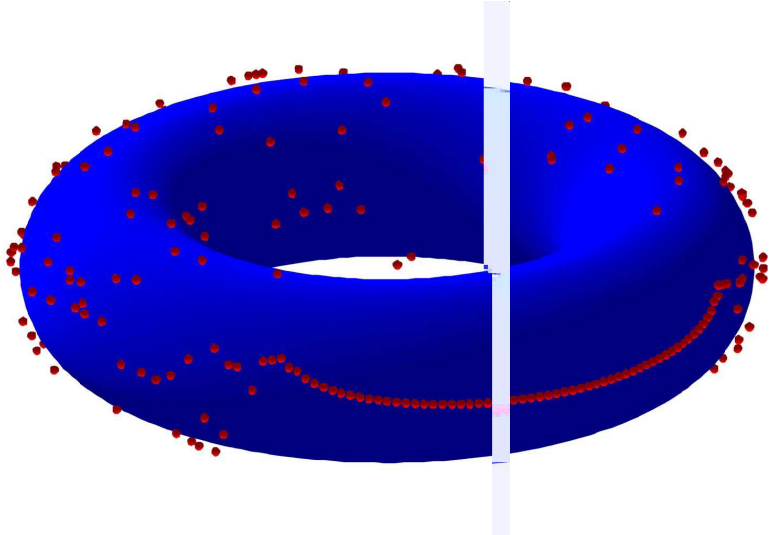}    }
    \caption{Phase pattern for a typical chimera state shown on the torus.
    Azimuthal angle indicates spatial position $x$.  Phase $\phi$ is
    constant along lines of latitude; the outer equator of the torus
    corresponds to $\phi=0$.}
    \label{fig:phiplot2}
\end{figure}

In this snapshot of the instantaneous phases, two distinct regions are
conspicuous. The oscillators near $x=0$ (and equivalently, $x=1$) are
phase-locked. All of them move with the same constant frequency; in a frame
rotating at this frequency, they would all look frozen. The smoothness and
flatness of the graph of $\phi(x)$ in this region indicates that these
oscillators are coherent as well, i.e., they are nearly in phase.

Meanwhile, the scattered oscillators in the middle of
Fig.~\ref{fig:phiplot} are drifting, both with respect to each
other and with respect to the locked oscillators. Their motion is
strongly nonuniform. They slow down when they pass near the locked
pack---which is why the dots appear more densely clumped at this
phase---and then speed up as they lap it.

\subsection{Puzzles}

When we first learned about the chimera state by reading
Ref.~\cite{kur03}, we were amazed by it.  How could such a thing
even be possible?

In fact, a little thought showed that it was provably
\emph{impossible} in two special cases that had been studied
previously:
\begin{itemize}
    \item     Global coupling: Chimera states can't occur for Eq.~\eqref{eq:kurphase} with
$G(x) \equiv 1$ and any choice of $0 \le \alpha<\pi/2$, because a
Lyapunov function exists for this case, demonstrating that almost
all solutions are attracted to the in-phase oscillation \cite{
watanabe93, watanabe94}.
    \item Sine coupling:  If $\alpha=0$, corresponding to a
pure sine coupling in  Eq.~\eqref{eq:kurphase}, chimera states are
impossible for any even kernel $G$ of any range.  This follows
because Eq.~\eqref{eq:kurphase} becomes a gradient system in the
frame rotating at frequency $\omega$.  Hence all attractors must
be fixed points, corresponding to phase-locked solutions in the
original frame, thus ruling out the possibility of coexisting
drift.
\end{itemize}
So the coexistence phenomenon must somehow rely on a conspiracy
between $\alpha \ne 0$ and the non-global nature of the coupling.
But how, exactly?

And for that matter, is the chimera state born as soon as $\alpha
\ne 0$, or at some value of $\alpha$ bounded away from zero? In
dynamical simulations like that shown in Fig.~\ref{fig:phiplot},
stable chimera states are observed only when $\alpha$ is close to,
but slightly less than, $\pi/2$.  Does that mean that these states
don't exist for smaller $\alpha$, or is just that their basins of
attraction shrink as $\alpha$ decreases?

Furthermore, what is the genealogy of the chimera state, in the
sense of bifurcation theory?  Is it born out of the vacuum, as a
pair of stable and unstable versions of itself?  Or does it emerge
when some more familiar attractor loses stability? For instance,
does it bifurcate off the fully incoherent state, in which
oscillators are uniformly scattered and drifting around the circle
at every $x$? That seems unlikely, since the phase pattern shown
in Fig.~\ref{fig:oldGdynamics} looks pretty far from total
incoherence; even its drifting oscillators show some clumping in
phase. So maybe the chimera state branches off the uniform
in-phase state?  But how can it, given that the in-phase state is
linearly stable for all $|\alpha| < \frac{\pi}{2}$?

Motivated by these puzzles, we have tried to understand where the
chimera state comes from and to pinpoint the conditions that allow
it to exist.  A brief report of our findings appeared in
\cite{us04}.

\subsection{Broader significance}
Aside from the questions it raises, we believe the chimera state is
also more broadly significant for nonlinear science, for two reasons.

First, it exemplifies the surprises that lurk in nonlocally coupled systems.
As Kuramoto and his colleagues have pointed out \cite{kur02,kur03,shima04,tanaka03,battogtokh00,kuramoto96,kuramoto95}, nonlocal coupling is a relatively
dark corner of nonlinear science in general, and nonlinear oscillator theory
in particular. Most previous work on coupled oscillators has focused on local
coupling, where the interactions are assumed to be solely between nearest
neighbors, or global coupling, where each oscillator interacts equally
strongly with all the others.  The intermediate case of nonlocal coupling is
natural to explore next, and has already revealed some interesting new forms
of dynamical behavior \cite{kuramoto96,kuramoto95}.

From a more applied perspective, nonlocal coupling is important to
investigate because it arises in diverse systems throughout physics,
chemistry, and biology.  Examples include Josephson junction arrays
\cite{phillips93}, chemical oscillators \cite{kur84},
epidemiological models of disease spread \cite{medlock03}, and the neural
networks underlying the patterns on mollusc shells \cite{murraybook,
ermentrout86}, localized neural ``bump" states
\cite{bumpstate1,bumpstate2,bumpstate3}, and ocular dominance stripes in the
visual cortex \cite{murraybook,swindale80}.

Second, the chimera state is by no means an oddity restricted to
Eq.~\eqref{eq:kurphase}.  On the contrary, it was first seen in
simulations of the complex Ginzburg-Landau equation with nonlocal
coupling \cite{kur02,kur03}, a fundamental model in the study of
pattern formation.  That equation in turn can be systematically
derived from a wide class of reaction-diffusion equations, under
particular assumptions on the local kinetics and diffusion
strength that render the effective coupling nonlocal
\cite{kur02,kur03,shima04,tanaka03}. Under an additional
assumption that the coupling is also sufficiently weak (in a
precise sense), Shima and Kuramoto \cite{shima04} show that the
original reaction-diffusion system can be further reduced to a
phase equation of the universal form \begin{displaymath}
   \frac{\partial \phi}{\partial t} = \omega -
   \int{d\br' G\paren{\br-\br'}\sin \bracket{\phi(\br) - \phi(\br') + \alpha}}
\end{displaymath}
where $\br$ labels the position of the oscillators and the kernel $G$ decays
exponentially with distance: $G(\br-\br') \propto
\exp\paren{-\kappa|\br-\br'|}$ . But this is just Eq.~\eqref{eq:kurphase}, if
the space is one-dimensional.  So there is good reason to expect that the
coexistence phenomenon will have some generality.

For example, in two dimensions, the coexistence of locked and
drifting oscillators manifests itself as an unprecedented kind of
spiral wave: one without a phase singularity at its center
\cite{kur03, shima04}. Instead, the oscillators in the core are
found to be completely desynchronized from each other and from the
uniform rotation of the spiral arms. In effect, the core
oscillators mimic a phase singularity by being incoherent.  A
better understanding of the one-dimensional case might shed light
on this remarkable new form of pattern formation.

\section{Summary of prior results} \label{sec:priorwork}

We begin by reviewing the results of Kuramoto and Battogtokh
\cite{kur02,kur03}. After uncovering the chimera state in their
simulations of Eq.~\eqref{eq:kurphase}, they were able to explain
much of its structure analytically.  Their elegant approach is a
generalization of Kuramoto's self-consistency argument for
globally coupled oscillators \cite{kur84, strogatz00}.

In this approach, one first transforms \eqref{eq:kurphase} by seeking a
rotating reference frame in which the dynamics become as simple as possible.
Let $\Omega$ denote the angular frequency of this rotating frame (to be
determined later, in the course of solving the problem), and let
\[
    \theta = \phi - \Omega t
\]
denote the phase of an oscillator relative to this frame. Next, introduce a
complex order parameter $R e^{i\Theta}$ that depends on space and time:
\begin{equation} \label{eq:orderparam}
    R(x,t) e^{i \Theta(x,t)} = \int_{0}^{1}{G\paren{x-x'} e^{i\theta(x',t)} dx'
        }~.
\end{equation}

To see what this order parameter means intuitively, note that the integral on
the right hand side of \eqref{eq:orderparam} performs a running average of
$e^{i \theta}$ over a window centered at $x$, with a width determined by the
width of the kernel $G$. Thus $0 \le R(x,r) \le 1$ can be viewed as a measure
of the local phase coherence at $x$, and $\Theta(x,t)$ represents the local
average phase. These two average quantities provide macroscopic proxies for
the overall state of the continuum of oscillators.

The real virtue of introducing the order parameter, however, is that we can
now rewrite the governing equation \eqref{eq:kurphase} as
\begin{equation} \label{eq:newkurphase}
    \frac{\partial\theta}{\partial t} = \omega - \Omega - R\sin \bracket{\theta -
        \Theta + \alpha}~,
\end{equation}
which makes it look as if the oscillators have decoupled, though
of course they are still interacting through $R$ and $\Theta$, to
which they each contribute through \eqref{eq:orderparam}.  This
observation suggests that the problem can be attacked by the
self-consistency arguments of mean-field theory, even though it is
not globally coupled.

Now comes the key step.  Suppose we restrict attention to
\emph{stationary} solutions, in which $R$ and $\Theta$ depend on
space but not on time. Now the equations truly do decouple, in the
following sense.  One can easily solve for the motion of the
oscillator at each $x$, subject to the assumed time-independent
values of $R(x)$ and $\Theta(x)$.  The oscillators with $R(x) \ge
|\omega-\Omega|$ asymptotically approach a stable fixed point
$\theta^*$, defined implicitly by
\begin{equation}
    \omega - \Omega = R(x)\sin \bracket{\theta^* - \Theta(x) + \alpha}
\end{equation}
The fact that they approach a fixed point in the rotating frame
implies that they are phase-locked at frequency $\Omega$ in the
original frame.  On the other hand, the oscillators with $R(x) <
|\omega - \Omega|$ drift around the phase circle monotonically. To
be consistent with the assumed stationarity of the solution, these
oscillators must distribute themselves according to an invariant
probability density $\rho(\theta)$. (To ease the notation here and
elsewhere, we have suppressed the dependence on $x$ whenever it's
clear from context.)  And for the density to be invariant, the
probability of finding an oscillator near a given value of
$\theta$ must be inversely proportional to the velocity there.
From \eqref{eq:newkurphase}, this condition becomes
\begin{equation} \label{eq:rho}
    \rho(\theta) = \frac{\sqrt{(\omega-\Omega)^2 - R^2}}{2\pi |\omega - \Omega - R \sin(\theta - \Theta + \alpha)|}
\end{equation}
where the normalization constant has been chosen such that
$\int_{-\pi}^{\pi}{\rho(\theta)d\theta} = 1$.

The resulting motions of both the locked and drifting oscillators must be
consistent with the assumed time-independent values for $R(x)$ and
$\Theta(x)$. To calculate the contribution that the locked oscillators make
to the order parameter \eqref{eq:orderparam}, observe that
\begin{eqnarray} \label{eq:sincos}
  \sin(\theta^* - \Theta + \alpha) &=& \frac{\omega-\Omega}{R} \nonumber \\
  \cos(\theta^* - \Theta + \alpha) &=& \pm \frac{\sqrt{R^2-(\omega-\Omega)^2}}{R}
\end{eqnarray}
for any fixed point of \eqref{eq:newkurphase}.  One can check that the
\emph{stable} fixed point of \eqref{eq:newkurphase} corresponds to the plus
sign in \eqref{eq:sincos}.  Hence
\begin{equation} \label{eq:exp}
  \exp\bracket{i(\theta^* - \Theta + \alpha)}   =  \frac{\sqrt{R^2-(\omega-\Omega)^2}+i(\omega-\Omega)}{R}
\end{equation}
which implies that the locked oscillators contribute
\begin{equation} \label{eq:lockedpart}
  \int dx' G(x-x') \exp \bracket{i\theta^*(x')} = e^{-i\alpha}
    \int dx' G(x-x') \exp \bracket{i\Theta(x')}
      \frac{\sqrt{R^2-(\omega-\Omega)^2}+i(\omega-\Omega)}{R}
\end{equation}
to the order parameter \eqref{eq:orderparam}.  Here the integral
is taken over the portion of the domain where $R(x') \ge
|\omega-\Omega|$.

Next, to calculate the contribution from the drifting oscillators, Kuramoto
and Battogtokh \cite{kur02,kur03} replace $\exp\bracket{i\theta(x')}$ in
\eqref{eq:orderparam} with its statistical average $\int_{-\pi}^{\pi} \exp(i
\theta) \rho(\theta) d\theta$.  Using \eqref{eq:rho} and contour integration,
they obtain
\[
  \int_{-\pi}^{\pi} \exp(i \theta) \rho(\theta) d\theta = \frac{i}{R}
    \paren{\omega-\Omega-\sqrt{(\omega-\Omega)^2-R^2}}~.
\]

Therefore the contribution of the drifting oscillators to the order parameter
is
\[
  \int dx' G(x-x') \int_{-\pi}^{\pi} \exp(i \theta) \rho(\theta) d\theta  =   i e^{-i\alpha} \int dx' G(x-x') \exp\bracket{i\Theta(x')}
    \frac{\omega-\Omega-\sqrt{(\omega-\Omega)^2-R^2(x')}}{R(x')}
\]
where now the integral is over the complementary portion of the domain where
$R(x') < |\omega-\Omega|$.

Notice something curious: the integrand on the right hand side of
the drifting contribution is exactly the same as that found
earlier in \eqref{eq:lockedpart} for the locked contribution; only
their domains differ. (This coincidence is not mentioned in
\cite{kur02,kur03}.)  To see that the two expressions agree, note
that
\[
    \sqrt{R^2-(\omega-\Omega)^2} + i(\omega-\Omega)    =     i \paren{\omega-\Omega-\sqrt{(\omega-\Omega)^2-R^2}}
\]
as long as we choose the branch corresponding to the ``$+i$" square root of a
negative number.

Hence the two contributions can be combined into a single integral, yielding
a slightly more compact version of the self-consistency equation derived in
\cite{kur02}:
\[
    R(x)\exp\bracket{i\Theta(x)} = i e^{-i\alpha} \int_0^1 G(x-x')
        \exp\bracket{i\Theta(x')}
        \frac{\omega-\Omega-\sqrt{(\omega-\Omega)^2-R^2(x')}}{R(x')} dx'~.
\]

To ease the notation a bit more, let
\begin{eqnarray} \label{eq:changenotation}
  \beta  &=& \frac{\pi}{2} - \alpha \nonumber \\
  \Delta &=& \omega - \Omega~.
\end{eqnarray}

Then the self-consistency equation becomes
\begin{equation} \label{eq:SCeqn}
 R(x) e^{i \Theta(x)} = e^{i \beta} \int_0^1{G(x-x') e^{i
 \Theta(x')}} {\frac{\Delta - \sqrt{\Delta^2 - R^2(x')}}{R(x')}} dx'~.
\end{equation}

Equation \eqref{eq:SCeqn} is to be solved for three unknowns---the
real-valued functions $R(x)$ and $\Theta(x)$ and the real number
$\Delta$---in terms of the assumed choices of $\beta$ and the
kernel $G(x)$. Notice that although $\omega$ itself is arbitrary
up to a constant, and hence so is $\Omega$, their difference
$\omega-\Omega$ is physically meaningful; it is determined by the
condition that the long-term dynamics become stationary in the
frame rotating at frequency $\Omega$.

Initially, we couldn't see how to solve the self-consistency
equation \eqref{eq:SCeqn} numerically.  We wrote to Kuramoto for
advice, and he described an iterative scheme to determine the
functions $R(x)$ and $\Theta(x)$, based on initial guesses
obtained from the dynamical simulations. The idea behind the
scheme is that the current estimates of $R(x)$ and $\Theta(x)$ can
be entered into the right hand side of \eqref{eq:SCeqn}, and used
to generate the new estimates appearing on the left hand side.

Still, that leaves open the question of how to determine $\Delta$.
We seem to have only two equations (given by the real and
imaginary parts of Eq.~\eqref{eq:SCeqn}) for three unknowns.
Fortunately, a third equation can be imposed to close the system.
Because \eqref{eq:SCeqn} is left unchanged by any rigid rotation
$\Theta(x) \to \Theta(x) + \Theta_0$, we can specify the value of
$\Theta(x)$ at any point $x$ we like; this freedom is tantamount
to choosing an origin in the rotating frame. A natural choice
would be to demand $\Theta(0)=0$, but as we'll see in Section
\ref{sec:exactsol}, another choice turns out to be more
convenient.

Kuramoto and Battogtokh \cite{kur02,kur03} confirm that the
self-consistency approach works: their results from numerical
integration of the dynamical equations \eqref{eq:kurphase} match
those obtained by solving the self-consistency equation
\eqref{eq:SCeqn} iteratively.

\begin{figure}[t] 
    \centering{
        \includegraphics[height=7cm]{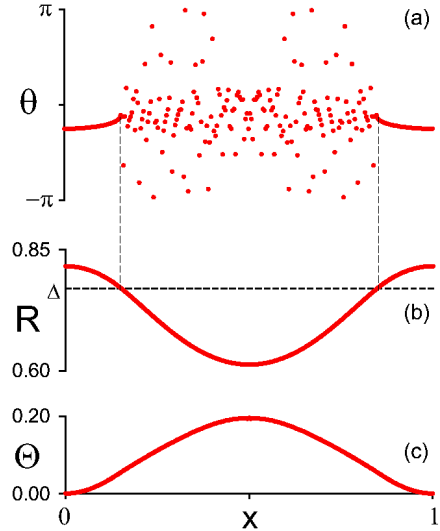}
    }
    \caption{Chimera state and order parameter curves for the exponential kernel
        $G(x)\propto \exp \paren{-\kappa |x|}$, as used by Kuramoto and Battogtokh \cite{kur02,kur03}.
        Parameters are the same as those in
        Fig.~\ref{fig:phiplot}.
        (a) Phase pattern for chimera state.
        (b) Local phase coherence $R(x)$, computed from \eqref{eq:orderparam}.
        Locked oscillators satisfy $R(x) \ge \Delta$.
        (c) Local average phase $\Theta(x)$, computed from
        \eqref{eq:orderparam}.}
    \label{fig:oldGdynamics}
\end{figure}

Figure \ref{fig:oldGdynamics} shows the chimera state along with
the graphs of $R(x)$ and $\Theta(x)$ for the parameters used in
Fig.~\ref{fig:phiplot}. The curves in
Fig.~\ref{fig:oldGdynamics}(b) and \ref{fig:oldGdynamics}(c) are
periodic and reflection-symmetric. In fact, they resemble cosine
waves, which made us wonder whether \eqref{eq:SCeqn} might have a
simple closed-form solution, perhaps in some perturbative limit as
a parameter tends to zero. To see where such a limit might come
into play, we hoped to first replicate the simulations of Kuramoto
and Battogtokh \cite{kur02,kur03} and then to explore parameter
space more widely.

\section{A first round of simulations} \label{sec:simulation}

Unfortunately, we couldn't find the chimera state in our early
simulations of Eq.~\eqref{eq:kurphase}.  No matter how we started
the system, it always converged to the in-phase state.  In the
report that announced the chimera state, Kuramoto \cite{kur03}
does not give precise details of the initial condition he used. He
describes it as a ``suitable single-humped initial phase pattern"
\cite[p. 219]{kur03} which we incorrectly took to mean something
like $\phi(x,0) = a + b \cos x$ or $e^{-a \cos x}$.

Eventually, we asked Kuramoto for help (again!), and he kindly
explained what he meant. (He also sent us his paper \cite{kur02}
with Battogtokh, where the description of the initial condition is
more explicit.) At each $x$, a uniform random number $\phi(x,0)$
is chosen within some interval whose width varies with $x$ in a
single-humped fashion. Specifically, the width is narrowest near
$x=0$ (mod 1), meaning that the oscillators are most nearly in
phase there, initially.  As $x$ increases toward the diametrically
opposite point of the domain at $x=\frac{1}{2}$, the phases are
scattered progressively over larger and larger regions on the
phase circle (meaning the oscillators are placed more and more
incoherently there, initially).  The effect of this procedure is
to give the system a jump-start, by placing it in a partially
coherent/partially incoherent state to begin with.

To be more precise, Kuramoto used a random distribution with a
Gaussian envelope: $\phi(x,0) = 6
\exp\bracket{-30\paren{x-\frac{1}{2}}^2} r(x)$, where $r(x)$ is a
uniform random number on the interval $-\frac{1}{2} \le r \le
\frac{1}{2}$. For the parameters used in
Fig.~\ref{fig:oldGdynamics}, this initial condition indeed evolves
to the chimera state reported in \cite{kur02, kur03}.

Then we ran simulations to see how far this state could be continued by
decreasing $\alpha$, knowing that it would have to disappear or lose
stability somewhere before $\alpha=0$. To track its fate along the way, we
also computed several statistics:
\begin{enumerate}
    \item the spatial average of $R(x)$, given by $\avg{R} = \int_0^1
        R(x)dx$;
    \item the amplitude of $R(x)$, defined as $R_{amp} = R_{max} - R_{min}$;
    \item $f_{\textrm{drift}}$, the fraction of the spatial domain occupied by
        drifting oscillators;
    \item the difference $\Delta = \omega - \Omega$ between the nominal
        frequency of individual oscillators and their collective frequency when
        locked; and
    \item $\overline{\Delta}_{\textrm{max}}$, the largest value of the time-averaged
        drift velocity relative to the rotating frame.  This quantity measures
        the average speed of the fastest drifting oscillator.  From
        \eqref{eq:newkurphase}, it can be calculated as
        $\max_x |\overline{\Delta}(x)| = \max_x |\sqrt{\Delta^2-R^2(x)}|$, where
        the maximum is taken only over the drifting oscillators.
\end{enumerate}

\begin{figure}[t] 
    \centering{    \includegraphics[height=7cm]{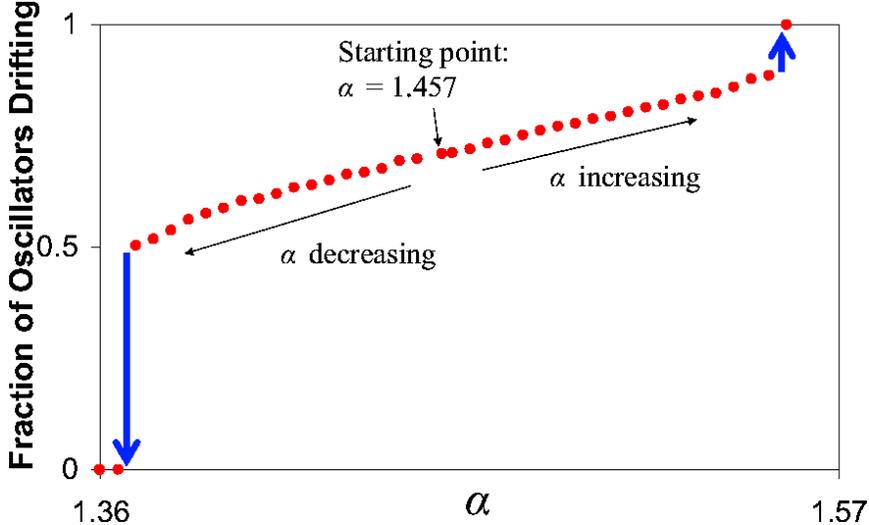}    }
    \caption{
        The fraction of oscillators drifting as the coupling parameter $\alpha$
        varies.  Here $\kappa=4.0$, $N = 256$ oscillators, $dt=0.025$ for 100,000
        iterations.}
    \label{fig:kappaconstant}
\end{figure}

Figure \ref{fig:kappaconstant} shows how $f_{\textrm{drift}}$
varies when $\kappa$ is held constant but $\alpha$ is changed
smoothly. We generated similar graphs for each of the statistics
mentioned above, and all showed a jump to the uniform synchronized
state as $\alpha$ decreased below some critical value $\alpha_c$.
From these results it appeared that when $\kappa = 4.0$, the
chimera state ceased to exist somewhere around $\alpha_c \approx
1.37$. The transition seemed to be discontinuous, which suggested
that $\alpha_c$ couldn't be calculated by a naive perturbation
expansion. If it was to be calculable at all, something more
subtle would be required.

The next step was to investigate how these results vary with
$\kappa$. Recall that the kernel in \eqref{eq:kurphase} is $G(x) =
C  \exp(-\kappa|x|)$, so $1 / \kappa$ sets a characteristic length
scale. Roughly speaking, it is the distance over which the
nonlocal coupling is substantial. So the limit $\kappa \to 0^+$
corresponds to global coupling $G(x) \equiv 1$. This can also be
checked directly, noting that the normalization constant for the
exponential kernel on the circle is given by $C = \frac{\kappa}
{2} (1-e^{-\kappa/2})^{-1}$.

\begin{figure}[t] 
    \centering{    \includegraphics[height=7cm]{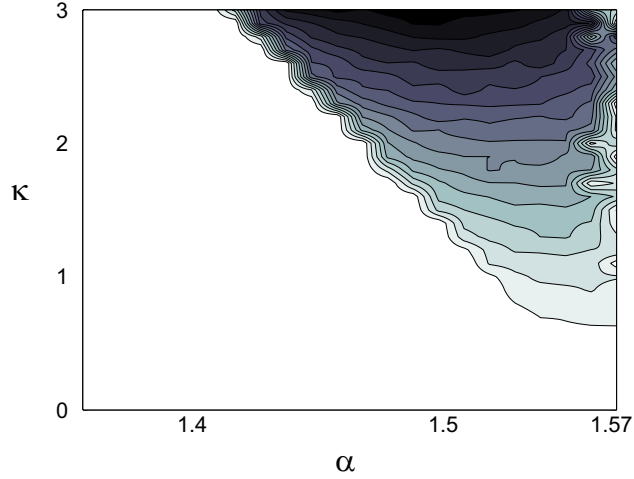}    }
    \caption{
        Amplitude of the curve $R(x)$, depicted as a contour plot in parameter space,
        and calculated by averaging over the instantaneous $R$ curves
        during numerical
        integration.  Here $G(x) \propto
        \exp(-\kappa|x|)$, $N = 80$ oscillators, the integration time step is $dt=0.025$,
        and integration continued for 20,000 iterations. Lighter colors indicate
        smaller amplitude; lightest is $R_{amp}=0.0$ and darkest is $R_{amp}=0.13$.}
    \label{fig:Ramp}
\end{figure}

Figure \ref{fig:Ramp} shows a rough contour plot of $R_{amp}$ in
the $(\alpha, \kappa)$ parameter plane.  Crude as this plot is,
its  message is still clear.  The stable chimera state evidently
lives in a wedge in parameter space, bounded on one side by the
line $\alpha = \frac{\pi}{2}$ and on the other by a curve
$\alpha=\alpha_c(\kappa)$ that is nearly a straight line. By its
very shape, the picture directs our attention to the corner of the
wedge, to the simultaneous limit as $\alpha \to \frac{\pi}{2}$
from below and $\kappa \to 0$ from above. Apparently something
crucial happens in that corner---the chimera state is born there.
And so this is where perturbation theory should be conducted.

\begin{figure}[t] 
    \centering{    \includegraphics[height=7cm]{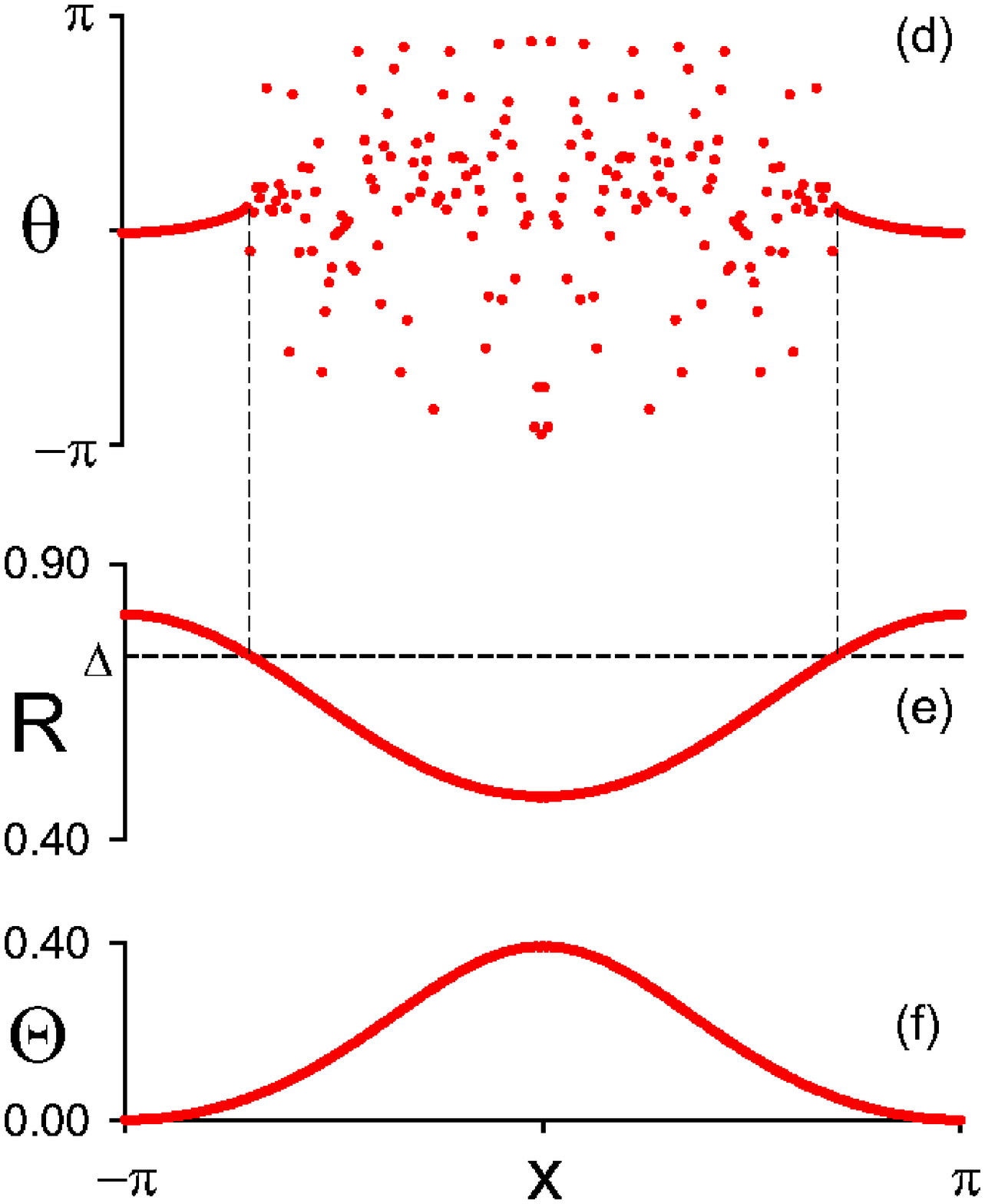}    }
    \caption{Chimera state and corresponding order parameter
    curves for the cosine kernel, shown in the same format as
    Fig.~\ref{fig:oldGdynamics}, and qualitatively similar to it.  Parameters are  $A=0.995$,
        $\beta=0.18$, $N=256$ oscillators; equation \eqref{eq:kurphase} was
        integrated with fixed time step $dt=0.025$ for 200,000 iterations,
        starting from $\phi(x)=6r\exp(-0.76x^2)$, where $r$ is a uniform
        random variable on $\bracket{-\frac{1}{2}, \frac{1}{2}}$.
    }
    \label{fig:newGdynamics}
\end{figure}

\begin{figure}[t] 
    \centering{  \includegraphics[height=7cm]{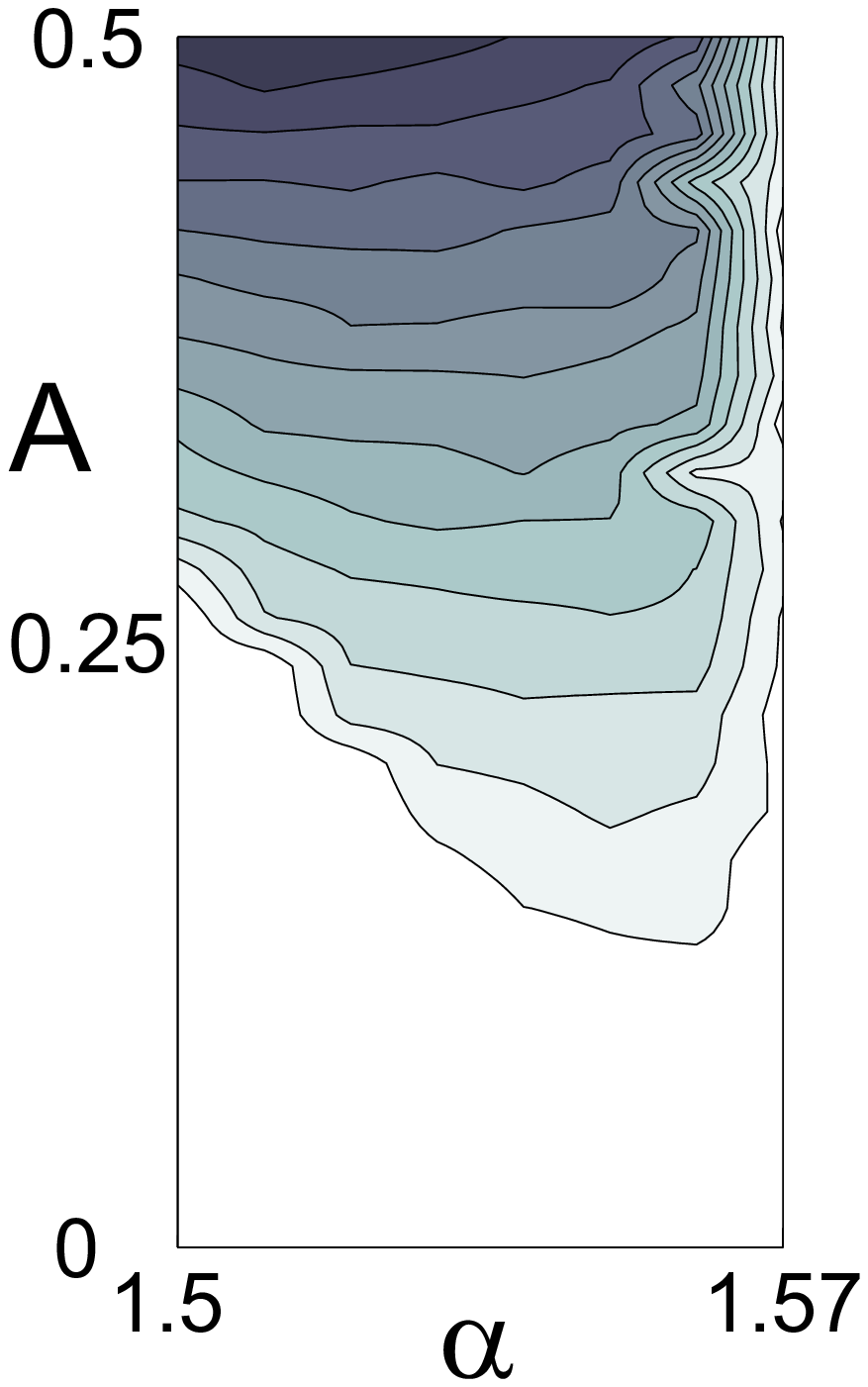}    }
    \caption{
        Contour plot of $R_{amp}$ for the chimera state with cosine kernel.
        Note the similarity to Fig.~\ref{fig:Ramp} for the exponential kernel.
        Here $G(x) = \frac{1}{2\pi}(1 + A\cos x)$, $N = 80$ oscillators,
        the integration time step was $dt=0.025$, and integration continued for 30,000
        iterations.  Same color scale as Fig.~\ref{fig:Ramp}.}
    \label{fig:RampnewG}
\end{figure}

To check that the wedge of Fig.~\ref{fig:Ramp} was not an artifact
of the exponential kernel assumed above, we also calculated the
corresponding contour plots for the cosine kernel \begin{equation}
\label{eq:newG}
    G(x) = \frac{1}{2\pi}(1 + A\cos x)~,
\end{equation}
where $0 \le A \le 1$.  Here the spatial domain has been redefined
to $-\pi \le x \le \pi$ for convenience, and to bring out its ring
geometry and the reflection symmetry of the chimera state. Figure
\ref{fig:newGdynamics} confirms that the cosine kernel gives a
similar chimera state to that for the exponential kernel used
above, while Fig.~\ref{fig:RampnewG} demonstrates that the wedge
in parameter space is preserved as well. All that is reassuring,
because as it happens, the cosine kernel also has the pleasant
property that it allows the self-consistency equation to be solved
analytically.

\section{An exactly solvable case} \label{sec:exactsol}

From now on, let $G(x)$ be given by the cosine kernel
\eqref{eq:newG}, and let the spatial domain be $-\pi \le x \le
\pi$ with periodic boundary conditions. For this case, we'll show
that the functional form of the order parameter can be obtained
exactly, which in turn yields the explicit $x$-dependence of
$R(x)$ and $\Theta(x)$. All the resulting expressions, however,
still contain two unknown coefficients, one real and the other
complex, which need to be determined self-consistently. In this
way, the two unknown functions in the self-consistency equation
are exchanged for two unknown numbers---a drastic reduction in the
difficulty of the problem.

The self-consistency equation \eqref{eq:SCeqn} is
\begin{equation} \label{eq:newSCeqn}
    R(x) e^{i \Theta(x)} = e^{i \beta} \int_{-\pi}^{\pi}{G(x-x') e^{i
        \Theta(x')}} h(x') dx'
\end{equation}
where we've introduced the notation
\begin{equation} \label{eq:hdef}
    h(x') = \frac{\Delta - \sqrt{\Delta^2 - R^2(x')}}{R(x')}~.
\end{equation}
Let angular brackets denote a spatial average:
\[
    \avg{f} = \frac{1}{2\pi} \int_{-\pi}^{\pi} f(x') dx'.
\]
Then, substituting the cosine kernel \eqref{eq:newG} into
\eqref{eq:newSCeqn} and expanding $G(x-x')$ by a trigonometric
identity, we find
\begin{eqnarray} \label{eq:trigID}
    R e^{i\Theta} &=& \frac{e^{i\beta}}{2\pi} \int_{-\pi}^{\pi}
        [1 + A \cos x \cos x' + A \sin x \sin x'] h(x') e^{i\Theta(x')} dx' \nonumber \\
    &=& e^{i\beta}\avg{he^{i\Theta}} + e^{i\beta}A\avg{he^{i\Theta}\cos x'}\cos x +
        e^{i\beta}A\avg{he^{i\Theta}\sin x'}\sin x \nonumber \\
    &=& c + a \cos x
\end{eqnarray}
where the coefficients $c$ and $a$ must satisfy their own version
of the self-consistency equations, now given by
\begin{equation}
\label{eq:c}
    c = e^{i\beta} \avg{h e^{i\Theta}}
\end{equation}
and
\begin{equation} \label{eq:a}
    a = Ae^{i\beta} \avg{h e^{i\Theta} \cos x'}~.
\end{equation}

Note that the coefficient of $\sin x$ vanishes in
\eqref{eq:trigID}.  This follows from the assumption that $R(x') =
R(-x')$ and $\Theta(x') = \Theta(-x')$, as suggested by the
simulations; then $h(x')$ in \eqref{eq:hdef} is also even, and so
the integral $\avg{he^{i\Theta}\sin x'}$ in \eqref{eq:trigID}
vanishes by oddness.  As we'll show next, this assumption of
reflection symmetry is self-consistent, in the sense that it
implies formulas for $R(x)$ and $\Theta(x)$ that indeed possess
this symmetry.

For example, to calculate $R(x)$ in terms of the unknown coefficients $a$ and
$c$, observe that
\begin{eqnarray} \label{eq:R2}
    R^2 &=& (R e^{i\Theta})(R e^{-i\Theta}) \nonumber \\
    &=& (c+a\cos x)(c^*+a^*\cos x) \nonumber \\
    &=& |c^2| + 2\re{ca^*}\cos x + |a|^2 \cos^2 x
\end{eqnarray}
which is an even function, and which also helps to explain why the
graph of $R(x)$ in Fig.~\ref{fig:oldGdynamics} resembled a cosine
wave.

Likewise, $\Theta(x)$ is an even function reminiscent of a cosine because
\begin{eqnarray} \label{eq:tanTheta}
    \tan \Theta(x) &=& \frac{R(x)\sin \Theta(x)}{R(x)\cos \Theta(x)} \nonumber \\
    &=& \frac{\im{c}+\im{a}\cos x}{\re{c}+\re{a}\cos x}~.
\end{eqnarray}

Another simplification is that $c$ can be taken to be purely real and
non-negative, without loss of generality.  This follows from the rotational
symmetry of the governing equations.  In particular, the self-consistency
equation \eqref{eq:newSCeqn} is left unchanged by any rigid rotation
$\Theta(x) \to \Theta(x) + \Theta_0$. Thus we are free to specify any value
of $\Theta(x)$ at whatever point we like. The most convenient choice is to
set
\[
    \Theta\paren{\frac{\pi}{2}} = 0~.
\]
Then at that value of $x$ the equation $R e^{i\Theta} = c + a\cos x$ reduces
to
\[
    R\paren{\frac{\pi}{2}} = c~.
\]
Since $R$ is real and non-negative, so is $c$.  Hence, we take
\begin{equation} \label{eq:imc}
    \im{c} = 0
\end{equation}
from now on.

The final step in closing the equations for $a$ and $c$ is to rewrite the
averages in \eqref{eq:c} and \eqref{eq:a} in terms of those variables. To do
so, we express $he^{i\Theta}$ as
\begin{eqnarray} \label{eq:hexp}
    he^{i\Theta} &=& \paren{Re^{i\Theta}}\frac{h}{R} \nonumber \\
    &=& (c+a\cos x) \frac{\Delta-\sqrt{\Delta^2-R^2(x)}}{R^2(x)} \nonumber \\
    &=& \frac{\Delta-\sqrt{\Delta^2-R^2(x)}}{c + a^* \cos x}
\end{eqnarray}
where we have used \eqref{eq:R2} and the real-valuedness of $c$ to simplify
the second line above.  Inserting \eqref{eq:R2} and \eqref{eq:hexp} into
\eqref{eq:c} and \eqref{eq:a}, we obtain the desired self-consistency
equations for $a$ and $c$:
\begin{equation} \label{eq:newc}
    c = e^{i \beta} \avg{ {\frac{\Delta-\paren{\Delta^2 - c^2 - 2 \re{a} c \cos x
       - |a|^2\cos^2 x}^{\frac{1}{2}}} {c + a^* \cos x}} }  \nonumber \\
\end{equation}
\begin{equation} \label{eq:newa}
    a = A e^{i \beta} \avg{ { \frac{\Delta-\paren{\Delta^2
       - c^2 - 2 \re{a} c \cos x - |a|^2\cos^2 x}^{\frac{1}{2}}} {c + a^* \cos x}
       \cos x} }~.
\end{equation}
This pair of complex equations is equivalent to four real
equations for the four real unknowns $c$, $\re{a}$, $\im{a}$, and
$\Delta$.  The solutions, if they exist, are to be expressed as
functions of the parameters $\beta$ and $A$.

\section{Clues based on numerics} \label{sec:numerics}
Before we plunge into the details of solving equations \eqref{eq:newc} and
\eqref{eq:newa} simultaneously, let's pause to remember what we're trying to
do.

We want to understand where the chimera state lives in parameter
space and how it bifurcates.  Guided by the simulations of Section
\ref{sec:simulation}, we expect that \eqref{eq:newc},
\eqref{eq:newa} should have chimera solutions throughout the
wedge-shaped region of parameter space shown in
Fig.~\ref{fig:RampnewG}. Assuming that's true, we hope that these
solutions will continue all the way down to the corner
$(\alpha,A)=(\frac{\pi}{2},0)$, corresponding to
$(\beta,A)=(0,0)$, where might be able to analyze them with
perturbation theory.

Our strategy, then, is to start by finding one solution to
\eqref{eq:newc}, \eqref{eq:newa}, by any means possible, for
parameter values anywhere in the wedge. Having found this
solution, we can use it as a base point for a numerical
continuation method. Then we proceed to dive into the corner,
following a straight line through parameter space between the base
point and the corner. In this way we convert the problem to a
one-parameter study of the solutions of \eqref{eq:newc},
\eqref{eq:newa}. Sufficiently close to the corner, we expect that
the solutions will display some sort of scaling behavior with
respect to the parameter.  That scaling will then suggest clues
about the right ansatz for a subsequent perturbation calculation.

So first we have to come up with a chimera solution to
\eqref{eq:newc}, \eqref{eq:newa}.  It's not just a matter of
plugging the equations into a standard root-finding package.  The
trouble is that these equations also have \emph{other} solutions
that we're less interested in, and we don't want the numerical
root-finding scheme to converge to them instead.

In particular, the in-phase solution, where all the oscillators
are locked at the same phase and none of them are drifting, has a
large basin of attraction that competes with that of the chimera
state.  To see what values of $a$, $c$, and $\Delta$ correspond to
the in-phase state, note that when $\phi(x,t)=\phi(x',t)$ for all
$x$ and $x'$, Eq.~\eqref{eq:kurphase} implies $\phi(x,t)=\phi_0 +
(\omega-\sin \alpha)t$.  Hence $R=1$ and therefore $c=1$ and
$a=0$. And because $\Omega=\omega-\sin \alpha = \omega - \cos
\beta$, we have $\Delta = \omega - \Omega = \cos \beta$. Thus
\begin{equation} \label{eq:inphase}
    (a, c, \Delta)_\textrm{in-phase} = (0, 1, \cos \beta)~.
\end{equation}
It's easy to check that this satisfies \eqref{eq:newc}, \eqref{eq:newa} for
all values of $A$ and $\beta$.

To reduce the chance that the root-finder will converge onto this
in-phase state, we need to concoct an initial guess that's very
close to a genuine chimera state.  To find one, we numerically
integrated Eq.~\eqref{eq:kurphase} using the cosine kernel, and
fit the resulting graphs of $R(x)$ and $\Theta(x)$ to the exact
formulas \eqref{eq:R2} and \eqref{eq:tanTheta}, to estimate the
values of $a$ and $c$.  The frequency difference $\Delta$ was
obtained directly from the simulation, by setting $\omega=0$ and
then computing $\Delta=\omega-\Omega=-\Omega$, where $\Omega$ is
observable as the collective frequency of the locked oscillators.

In this way we estimated $a=0.156-0.072i$, $c=0.591$,
$\Delta=0.720$ for the stable chimera state at parameter values
$A=0.99$, $\beta=0.081$. We fed this starting guess into the
Matlab root-finding and numerical continuation program MatCont
\cite{matcont} and found rapid convergence to $a=0.162-0.051i$,
$c=0.588$, $\Delta=0.723$. From there, we could continue the
solution in either $A$ or $\beta$ or some combination, as we saw
fit.

\begin{figure}[t] 
    \centering{  \includegraphics[height=5cm]{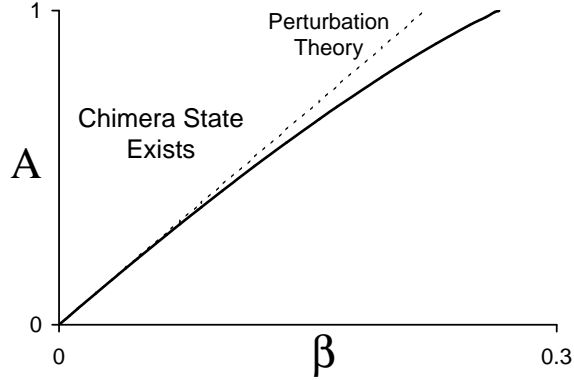}    }
    \caption{
        The region of parameter space in which the chimera state
        exists. Solid line, exact boundary determined by
        numerical solution of \eqref{eq:newc} and \eqref{eq:newa};
        dashed line, leading order approximation to
        this boundary obtained by perturbation theory
        (see text).
        }
    \label{fig:wedge}
\end{figure}

This approach enabled us to track the chimera state throughout
parameter space, until it disappeared along a critical curve
corresponding to the boundary of the wedge shown earlier. The
results of this calculation are shown in Fig.~\ref{fig:wedge}. As
expected, the boundary of the region is nearly a straight line,
and it extends down to the origin.

\section{Perturbation theory}
\begin{figure}[t] 
    \centering{  \includegraphics[height=6cm]{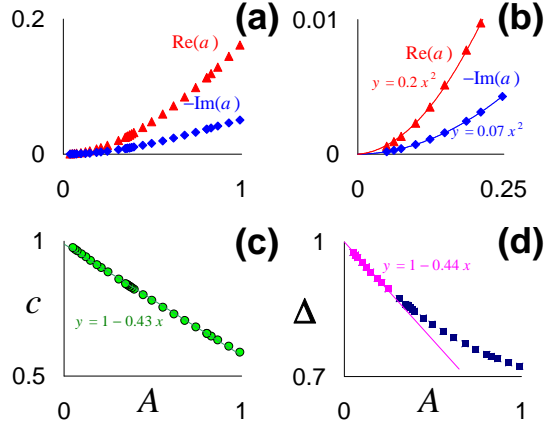}    }
    \caption{Scaling laws near the origin in parameter space, along the line $A = 12.375 \beta$.
        Data were collected from numerical continuation of a known chimera state,
        for an ensemble of parameter values.
        Approximate fits were then determined by least-square
        regression.
        (a) Scaling of real and imaginary
        parts of $a$; (b) Zoom of panel (a) near origin in parameter space.  Note that
        curves are quadratic; (c)
        Linear scaling of real-valued variable $c$; (d) Scaling of $\Delta$.
        Note that $\Delta$ scales linearly for small values of $A$ (purple).
        }
   \label{fig:scaling}
\end{figure}

The next step is to look for scaling laws to guide our
perturbation calculations. Figure \ref{fig:scaling} shows the
results of numerical continuation starting from $(\beta, A) =
(0.08, 0.99)$ and moving along the line $A=12.375\beta$ towards
the origin, all the while remaining within the wedge shown in
Fig.~\ref{fig:wedge}.  The observed behavior of the variables
along that line suggests the following ansatz near the origin:
\begin{eqnarray}\label{eq:ansatz}
   \Delta & \sim & 1 + \Delta_1 \epsilon + \Delta_2 \epsilon^2 \nonumber \\
   c & \sim & 1 + c_1 \epsilon + c_2 \epsilon^2 \nonumber \\
   \re{a} & \sim & u \epsilon^2 \nonumber \\
   \im{a} & \sim & v \epsilon^2
\end{eqnarray}
where we have introduced $\epsilon = A$ as the small parameter.

Next, we assume that this ansatz continues to hold along other
lines through the origin.  Such lines can be parametrized as
\begin{eqnarray*}
    A=\epsilon, \\
    \beta=\beta_1 \epsilon~,
\end{eqnarray*}
where $A$ and $\beta$ tend to zero simultaneously as $\epsilon
\rightarrow 0$. Here $\beta_1$ is a free parameter inversely
related to the slope of the lines. Thus the asymptotic shape of
the wedge in Fig.~\ref{fig:wedge}, sufficiently close to the
origin, will be determined from the maximum and minimum values of
$\beta_1$ for which a perturbative solution exists.

Substituting the ansatz \eqref{eq:ansatz} into the
self-consistency equation~\eqref{eq:newc} for $c$, and retaining
only terms up to $\order{\sqrt\epsilon}$ gives
\begin{eqnarray*} \label{eq:lowordernot}
    1 + \order{\epsilon} &=& (1+i\beta_1 \epsilon) \avg{ {\frac{1+\Delta_1 \epsilon -
        \paren{1+2\Delta_1 \epsilon - 1 - 2c_1 \epsilon}^{\frac{1}{2}}}
        {1+c_1 \epsilon}} }  \nonumber \\
    &=& 1 - \sqrt{2} \sqrt{\Delta_1-c_1}\sqrt{\epsilon} + \order{\epsilon}~,
\end{eqnarray*}
implying that
\begin{equation} \label{eq:loworder}
    \Delta_1 = c_1~.
\end{equation}

Now we retain terms up to $\order{\epsilon}$ on both sides, and
apply Eq.~\eqref{eq:loworder} whenever necessary to cancel terms.
At this order, Eq.~\eqref{eq:newc} becomes
\begin{equation} \label{eq:nextorder}
    1 + c_1 \epsilon = 1+i\beta_1 \epsilon - \epsilon \sqrt{2} \avg{
        \sqrt{ (\Delta_2-c_2) - u \cos x}}~.
\end{equation}
To simplify notation, let
\begin{equation}  \label{eq:deltadef}
    \delta = \Delta_2-c_2~.
\end{equation}
After breaking up the previous expression \eqref{eq:nextorder}
into two equations for the real and imaginary parts, and equating
terms of $\order{\epsilon}$, we get
\begin{eqnarray}
  \label{eq:c1eqn}
  c_1     &=& -\textrm{Re}{\bracket{\sqrt{2} \avg{\sqrt{\delta - u \cos x}}}} \\
  \label{eq:betaeqn}
  \beta_1 &=&  \textrm{Im}{\bracket{\sqrt{2} \avg{\sqrt{\delta - u \cos
  x}}}}~.
\end{eqnarray}
Repeating the same expansion to $\order{\epsilon}$ in the
self-consistency equation~\eqref{eq:newa} for $a$ yields two
analogous equations:
\begin{eqnarray}
  \label{eq:ueqn}
  u       &=& -\textrm{Re}{\bracket{\sqrt{2} \avg{\cos x \sqrt{\delta - u \cos x}}}}\\
  \label{eq:veqn}
  v       &=& -\textrm{Im}{\bracket{\sqrt{2} \avg{\cos x \sqrt{\delta - u\cos
      x}}}}~.
\end{eqnarray}

The equations \eqref{eq:c1eqn}--\eqref{eq:veqn} form a closed
system for the variables $(c_1, u, v, \delta)$, given the
parameter $\beta_1$.  But to solve these equations, it proves more
convenient to regard $\beta_1$ as a variable, and $\delta$ as a
parameter; we adopt this point of view in what follows.

There's another important structural aspect of equations
\eqref{eq:c1eqn}--\eqref{eq:veqn}, namely, that \eqref{eq:ueqn} is
distinguished in that it involves only two unknown quantities.  It
has the form $u = f(u,\delta)$ and can be solved numerically for a
given $\delta$. When a solution exists, all other variables
($c_1$, $\beta_1$, $v$) can be generated parametrically from the
($u$, $\delta$) pair.  Thus, the problem of solving equations
\eqref{eq:c1eqn}--\eqref{eq:veqn} reduces to a root-finding
exercise in one dimension instead of four.

\begin{figure}[t] 
    \centering{  \includegraphics[height=4cm]{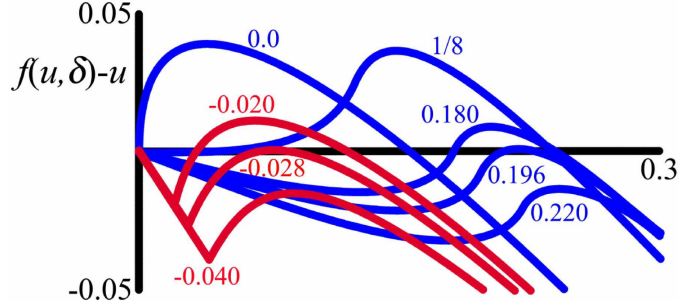}    }
    \caption{Roots of Eq.~\eqref{eq:ueqn} for various values of $\delta$.  Red
    indicates negative $\delta$ and blue positive $\delta$.  For $\delta<-0.028$ there are
    no roots; for $-0.028 < \delta < 0$, two roots; for
    $0 < \delta < \frac{1}{8}$, one root;  for
    $\frac{1}{8} < \delta < 0.196$, two roots; and for $\delta > 0.196$, no
    roots .
    }
    \label{fig:roots}
\end{figure}

Figure \ref{fig:roots} plots the graph of the difference
$f(u,\delta)-u$ for several values of $\delta$.  The zeros of this
graph correspond to the solutions of \eqref{eq:ueqn}, and yield
the desired ($u(\delta)$, $\delta$) pairs.  These are then
substituted into the remaining equations to obtain $c_1(\delta)$,
$\beta_1(\delta)$, and $v(\delta)$, from which various quantities
of physical interest can be derived.

\subsection{Calculation of $f_{\textrm{drift}}$}

For example, we can use the perturbative solution to find
$f_{\textrm{drift}}$, the fraction of the system that is drifting.
It is most convenient to calculate this quantity first in terms of
$\delta$, and then later re-express it in terms of the more
natural control parameter $\beta_1$.

To find the drifting oscillators, recall from
Fig.~\ref{fig:newGdynamics} that the cutoff between the locked
portion and the drifting portion occurs at the crossover value
$x=x_c$ where $R(x_c) = |\Delta|$. Substituting \eqref{eq:R2} for
$R^2$ and equating this to $\Delta^2$, we obtain \begin{equation}
\label{eq:xc}
    |c|^2 + 2\re{ca^*}\cos x_c +|a|^2 \cos^2 x_c =  \Delta^2~.  \nonumber
\end{equation}
Plugging in the ansatz \eqref{eq:ansatz} and keeping terms up to
order $\epsilon^2$, we find
\begin{equation}
    1 + 2c_1\epsilon + (c_1^2+2c_2)\epsilon^2 + 2u\epsilon^2 \cos x_c =
        1 + 2 \Delta_1 \epsilon + (\Delta_1^2+2\Delta_2) \epsilon^2~.  \nonumber \\
\end{equation}
Finally, because of \eqref{eq:loworder}, this simplifies to
\begin{equation} \label{eq:xc2}
    \cos x_c = \frac{\Delta_2-c_2}{u} = \frac{\delta}{u}~.
\end{equation}
Since the spatial domain of the ring has length $2 \pi$ and $2
x_c$ is the length of the region occupied by drifting oscillators,
the fraction of the chimera state corresponding to drifting
oscillators is $f_{\textrm{drift}} = x_c/\pi$, and hence
\begin{equation} \label{eq:fdrift}
 f_{\textrm{drift}} = \frac{1}{\pi} \abs{\cos^{-1} \paren{\frac{\delta}{u(\delta)}}}~.
\end{equation}

\begin{figure}[t] 
    \centering{  \includegraphics[height=4cm]{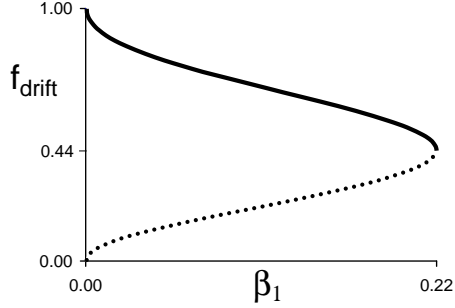}    }
    \caption{Fraction of chimera state consisting of drifting oscillators
    as a function of $\beta_1$. Solid line indicates stable chimera,
    dotted line indicates unstable.  The maximum $\beta_1$
    determines the line bounding the wedge-shaped existence region in Fig.~\ref{fig:wedge}.
    }
    \label{fig:fdrift}
\end{figure}

Figure \ref{fig:fdrift} plots the numerically computed
$f_{\textrm{drift}}(\delta)$ against $\beta_1(\delta)$. The curve
has a turning point at $\beta_1 \approx 0.22$, when about 44\% of
the system is drifting.  Presumably, this turning point stems from
a saddle-node bifurcation in the underlying dynamics.  In our
simulations, we only see the upper branch of this curve,
suggesting that this corresponds to the \emph{stable} version of
the chimera state. The reciprocal of the critical $\beta_1$ is
about 4.5, which is the slope of the dashed line shown in
Fig.~\ref{fig:wedge}, in excellent agreement with the boundary of
the wedge found numerically.

\begin{figure}[t] 
    \centering{  \includegraphics[height=8cm]{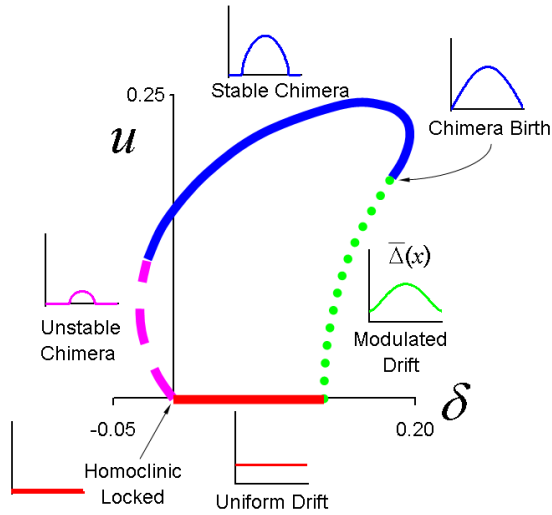}    }
    \caption{Diagram of bifurcations giving rise to the chimera state in
    $u-\delta$ plane.  Insets show average frequency ${\overline{\Delta}}$
    versus $x$.  Please see text for definitions of the perturbative
    variables $\delta$ and $u$, and for a detailed explanation of this
    figure.
    }
    \label{fig:udelta}
\end{figure}
\subsection{Birth and Death of the Chimera State} \label{sec:chimera_cycle}
Although the parametric dependence of $f_{\textrm{drift}}$ seems
to be conveniently expressed with respect to $\beta_1$, that
representation conceals a lot. Several dynamically distinct states
of the system are invisible because they are all squeezed onto a
single point $(\beta_1, f_{\textrm{drift}}) = (0,1)$, as we'll see
below.  It's much more revealing to use $\delta$ instead of
$\beta_1$.

Therefore, we now examine the system in the set of coordinates shown in
Fig.~\ref{fig:udelta}, with $u$ plotted vertically and $\delta$ horizontally.
This picture is a compendium of \emph{all} the stationary states of the
system---the stable and unstable chimera, along with other states that we
haven't mentioned yet, which we call uniform drift, modulated drift, and
homoclinic locked states.  The virtue of this representation is that it
allows us to see each bifurcation that occurs as the chimera state comes into
existence and later disappears. Beginning at the origin and moving
counterclockwise around the kidney-bean shaped cycle, we have:

\begin{enumerate}
    \item
        \emph{Homoclinic locked state:} $u=\delta=0$.
        Here, all the oscillators are locked in phase, and hence frozen in the rotating
        frame.  Accordingly, the average frequency $\overline{\Delta}(x)$ of the
        oscillators vanishes for all $x$, as shown in the
        inset.
        But one can show that this state is not linearly stable.
        In fact, the exact, non-perturbative counterpart of this state is the in-phase locked
        state \eqref{eq:inphase} at the critical parameter value $\beta=0$, where this
        state undergoes a homoclinic saddle-node bifurcation.

    \item
        \emph{Spatially uniform drift:} For $u=0$ and small $\delta > 0$, the system has a stationary state in
        which all the oscillators drift in a way that varies strongly in time but remains uniform in
        $x$. The order parameter $R(x)$ is independent of $x$ and close to 1, meaning
        that the oscillators are nearly in phase for nearly all of the time.
        An individual oscillator executes a jerky motion
        around its phase circle, lingering near
        $\theta =0$ and then whipping around the rest of the phase circle back to
        this point.  The associated plot of $\overline{\Delta}(x)$
        is flat because of the uniformity in $x$.

        In terms of the perturbative variables used in Fig.~\ref{fig:udelta}, this state appears on the line
        $u=0$ with $\delta > 0$.  Then \eqref{eq:veqn} shows that $v=0$ as well;
        hence $a = 0$, to $\order{\epsilon^2}$.  So \eqref{eq:R2} implies that $R(x)$ must be real and constant
        and \eqref{eq:tanTheta} implies that $\Theta(x)=0$.  Equation~\eqref{eq:betaeqn} tells us that
        such a state is possible only
        if $\beta_1=0$, which suggests that one can find an exact, non-perturbative version of the
        uniform drift state when $\beta=0$. Indeed, seeking a solution of
        the self-consistency equation \eqref{eq:newSCeqn}
        with $\beta=0, \Theta(x)=0$, and constant $R(x)$, one finds $R^2
        = \Delta-\sqrt{\Delta^2-R^2}$ since the kernel $G$ is normalized.
        Hence, along this line $\beta=0$, $\Theta(x)=0$, and $R(x) = R = \sqrt{2\Delta-1}$.
    \item
        \emph{Onset of spatial structure:} At the lower right corner of the kidney bean,
        the function $f(u,\delta)-u$ (Fig.~\ref{fig:roots})
        becomes tangent at the origin, introducing a new branch of solutions with
        $\Theta=0$ and $\beta=0$ but with the coherence $R$ varying spatially.
        This is the birth of spatial structure in the system.
        It happens for
        $\delta=\frac{1}{8}$.  The non-perturbative generalization of this result is
        $\Delta=2/(2+A)$.
        (See Appendix \ref{app:spatialstructure}.)
    \item
        \emph{Modulated drift:} Along the first curved branch, all oscillators continue to drift, but now there is
        spatial structure in the $R(x)$ curve, leading to a modulated pattern of average velocities (see inset).
        However, the average angle $\Theta(x)$ is still
        identically
        $0$.
    \item
        \emph{Chimera birth:} At the point where $u=\delta$, the first
        locked oscillators are born.
        For the first time, $v$ and $\beta_1$ become nonzero (see Eq.
        \eqref{eq:betaeqn},\eqref{eq:veqn}).  Until this point, all
        of the states have been confined to the vertical axis of
        Fig.~\ref{fig:wedge}; now we finally we move off the wall.  The
        curve of average velocities $\overline{\Delta}(x)$ touches
        the $x$-axis at a single point.  Meanwhile, the system develops spatial structure in its average phase:
         $\Theta(x)$ is no longer identically zero.

        This bifurcation can be shown to occur at $\delta=16/(9\pi^2)$,
        by evaluating the integral in \eqref{eq:ueqn} with $u=\delta$ ; also,
        see Appendix \ref{app:firstlocked} for an exact calculation of the chimera state at birth.
    \item
        \emph{Stable chimera:} Along the top of the kidney bean, the chimera state is dynamically
        stable.  After its birth from the spatially modulated drift
        state, it gradually develops an increasing fraction of locked oscillators
        as we move in the counterclockwise direction.  Locked
        oscillators correspond to the zero part of the $\overline{\Delta}(x)$
        curve (they appear motionless because the reference frame was chosen to co-rotate with
        them).
    \item
        \emph{Saddle-node bifurcation:} As we continue to move counterclockwise, the value of $\beta_1$
        grows (along with the fraction of locked oscillators), and reaches its maximum
        at the point where the stable
        chimera ceases to exist.  The disappearance is a result of a saddle-node
        bifurcation---a collision with an unstable chimera state---and occurs when about
         44\% of the system is drifting.
    \item
        \emph{Unstable chimera:} Along the unstable dashed branch, the fraction of locked oscillators continues
        to grow.  But the value of $\beta_1$ now begins to decrease, indicating a
        movement back towards the wall in Fig.~\ref{fig:wedge}.  The system returns to
        its original state when all oscillators become locked, with
        $\delta = u = \beta_1 = 0$ and $R$=1.
\end{enumerate}

Taking a step back, we can see an interesting message of
Fig.~\ref{fig:udelta}. The stable and unstable chimera states are
continuously connected though the branches of drifting states,
shown in solid red and dotted green lines.  If we had used the
$\beta_1$ representation instead (as in Fig.~\ref{fig:fdrift}),
both of these connecting branches would have shrivelled down to a
point.  The two kinds of chimera states would seem disconnected in
a way that they really aren't.

\section{Discussion}

Our main result in this paper is an exact solution for the chimera
state, for the special case of a cosine kernel.  That solution
also shed light on the bifurcations which create and destroy the
chimera.

In retrospect, it's not surprising that a cosine kernel would make
the self-consistency equation \eqref{eq:SCeqn} as tractable as
possible, because the right hand side of \eqref{eq:SCeqn} is a
convolution integral, and trigonometric functions behave nicely
under convolution.  For this reason, it should be straightforward
to extend the calculations to include more harmonics in $G$. Using
the same argument as in Section \ref{sec:exactsol}, one can see
that the exact solution for the order parameter \eqref{eq:trigID}
will have the same number of harmonics as $G$ has. This approach
would then give a systematic way to solve the self-consistency
equation for any kernel representable as a finite Fourier cosine
series.  By taking more and more terms, this approach also gives a
way to approximate results for any even kernel, as long as it is
representable by a Fourier series.

Unfortunately, the trick of choosing a special kernel may not work
as well in two (or three) spatial dimensions.  That could limit
the applicability to two-dimensional chimeras, such as the novel
spiral waves computed numerically in \cite{kur03}. Nevertheless,
the idea of seeking a tractable kernel that can simplify the
problem may itself be useful.

Another caveat is that, despite its usefulness as a mathematical
tool, the perturbative approach adopted here does not give a
rigorous understanding of the bifurcations in the original
problem.  One would like to understand the bifurcation scenarios
for \emph{all} values of the coupling parameter $A$, which
essentially measures how far the nonlocal coupling deviates from
strictly global coupling.  In Appendices
\ref{app:spatialstructure} and \ref{app:firstlocked}, we show two
results along these lines.

One interesting aspect of the perturbative approach is that it
draws our attention to the special parameter values $A=0$,
$\beta=0$ (or equivalently $\alpha = \pi/2$). Here the system has
global cosine coupling and is known to be completely integrable
\cite{watanabe93, watanabe94}.  So in a sense, what we have done
in this paper is perturb off this extremely degenerate system,
which raises the question of whether other, unforeseen attractors
might also lurk nearby, for different choices of initial
conditions.

The surprising nature of the chimera state makes us wonder if it
could be created artificially in a laboratory experiment, or
possibly even occur naturally in some system.

As a first attempt to judge whether this might be possible, we
tried to integrate the phase equation \eqref{eq:kurphase} with
slightly nonuniform frequencies $\omega_i$, to mimic the
inhomogeneities that would occur in any real system, and to test
whether the chimera is an artifact of assuming identical
oscillators. We added a uniform random variable $r \in [-B,B]$ to
the native frequency $\omega$ for each oscillator, and we found
the chimera state persisted, as long as $B$ was not too
large---less than about $4\%$ of $\Delta$ (the frequency
difference between the locked oscillators' $\Omega$ and the mean
natural frequency $\omega$). This estimate should be conservative
when compared with a presumably more realistic Gaussian random
distribution of $\omega_i$.

There are several possibilities for experimental systems where the
distinctive effects of nonlocal coupling, including the chimera
state, might be observed. Laser arrays seem to be good candidates.
In some cases, such as semiconductor arrays with evanescent
coupling \cite{winful88, li92}, they are governed by equations
similar to \eqref{eq:kurphase}, though these are usually
approximated as nearest-neighbor. Likewise, phase equations of
this form arise in the description of coupled electronic
phase-locked loops, and superconducting arrays of Josephson
junctions \cite{wiesenfeld98, swift92}.  Finally, an
idealized model of biochemical oscillators, coupled by a
diffusible substance that they all produce, can give rise to an
effectively nonlocal coupling and chimera states; indeed, this was
the motivating example that led Kuramoto and his colleagues to
their discovery.

Whether or not the chimera state turns out to experimentally
realizable, it is fascinating in its own right, as a strange new
mechanism for pattern formation in spatially extended nonlinear
systems.  Its existence underscores how much still remains to be
discovered, even in what would seem to be the simplest possible
model of pattern formation: a one-dimensional collection of
identical oscillators.

\section{acknowledgments}

\textbf{} Research supported in part by the National Science
Foundation. We thank Yoshiki Kuramoto for helpful correspondence,
Steve Vavasis for advice about solving the self-consistency
equation numerically, Bard Ermentrout for drawing our attention to
bump states in neural systems, Dan Wiley and Herbert Hui for
helpful discussions, and Kim Sneppen and the Niels Bohr Institute
for being such gracious hosts.

\newpage

\appendix
\section{Onset of Spatial Structure} \label{app:spatialstructure}
We now show that the birth of spatial structure can be calculated
non-perturbatively.  We have already seen in Section
\ref{sec:chimera_cycle} that when $\beta=0$,  the system has an
exact state of spatially uniform drift with constant coherence
$R(x) \equiv \sqrt{2\Delta-1}$ and average phase $\Theta(x) \equiv
0$. For this special state, the modulation amplitude $a=0$ and the
mean level of the coherence $c=R = \sqrt{2\Delta-1}$.  At the
bifurcation that creates spatial variation in the coherence, the
real part of $a$ becomes nonzero; at leading order in perturbation
theory, this bifurcation takes place at $\delta = \frac{1}{8}$.
Meanwhile, the imaginary part of $a$ remains zero, which means
that $\Theta(x) \equiv 0$ still holds.

To generalize this result to the non-perturbative case, we seek
conditions for a second branch of solutions to bifurcate off the
uniform drift state. Since $\Delta = (1+c^2)/2$ for the drift
state, we consider a slight perturbation \begin{equation}
    \Delta = (1+c^2)/2 + \eta~,
\end{equation}
where $\eta$ is a small deviation.  Also, since $a=0$ for the
uniform drift state, we may take $a$ itself as a small deviation.
Plugging all this into the self-consistency equation
\eqref{eq:newc} gives:
\begin{equation} \label{eq:blah}
    c = \avg{ {\frac{(1+c^2)/2 + \eta -\bracket{((1+c^2)/2 + \eta)^2
        - c^2 - 2 a c \cos x
        - a^2\cos^2 x}^{\frac{1}{2}}} {c + a \cos x}} }~.
\end{equation}
Now expand in a two-variable Taylor series for small $\eta$ and
$a$, and integrate over $x$ to obtain:
\begin{equation}
\label{eq:cseries}
    0 = \bracket{\frac{2c}{c^2-1}\eta + \order{\eta^2}}
        + \bracket{
            \frac{c(c^2-3)(c^2+1)}{2(1-c^2)^3}
            + \frac{c(c^6-5c^4+19c^2+9)}{(1-c^2)^5}\eta + \order{\eta^2}
        }a^2 + \order{a^4}~.
\end{equation}
Repeating the approach for the second self-consistency equation
\eqref{eq:newa} gives:
\begin{equation}  \label{eq:aseries}
    0 = \bracket{
            \frac{2(c^2-1)+A(c^2+1)}{2(c^2-1)}
            + \frac{A(c^4-4c^2-1)}{(c^2-1)^3}\eta
            + \order{\eta^2}}a
        + \order{a^3}~.
\end{equation}

To locate where another branch of solutions bifurcates off the
uniform drift solution, we inspect the linearization of the
algebraic system above, given by the Jacobian matrix
 \begin{equation} \label{eq:jacmatrix}
    \bracket{  \begin{array}{ccc}
        \frac{2c}{(c^2-1)^2} & 0 \\
        0 & \frac{2(c^2-1)+A(c^2+1)}{2(c^2-1)} \\
    \end{array}}
\end{equation}
If the determinant of the Jacobian is nonzero, the implicit
function theorem tells us that no other solutions exist nearby.
Hence, the existence of a continuously bifurcating branch requires
that the determinant vanish. Setting the determinant equal to zero
yields $c=0$ or $2(c^2-1)+A(c^2+1)=0$. Plugging in the value of
$c$ about which we're linearizing, $c = \sqrt{2\Delta-1}$, and
solving for $\Delta$ finally gives the bifurcation condition
\begin{equation} \label{eq:crit_Delta}
    \Delta_c = \frac{2}{A+2}~.
\end{equation}

To compare this with our earlier result from first-order
perturbation theory, we express the perturbative variable $\delta$
at this critical point by using its definition from
\eqref{eq:deltadef} above and the property in \eqref{eq:loworder}.
Since $\Delta - c = \delta \epsilon^2 = \delta A^2$ (ignoring
higher order terms), we have $\delta = \frac{\Delta-c}{A^2}$. So
\begin{eqnarray}
    \delta_c &=& \frac{\Delta_c-c}{A^2}   \nonumber \\
    &=& \frac{1}{A^2}\paren{\frac{2}{A+2}-\sqrt{2\Delta_c-1}}  \nonumber \\
    &=& \frac{1}{A^2}\paren{\frac{2}{A+2}-\sqrt{\frac{2-A}{2+A}}}  \nonumber \\
    &=& \frac{1}{8} - \frac{1}{16}A + \frac{5}{128}A^2 - \frac{5}{256}A^3 +
        \order{A^4}~,
\end{eqnarray}
which agrees with our perturbative prediction that $\delta_c=\frac{1}{8}$ in
the limit that $A \to 0$.

\newpage
\section{Birth of the Chimera State} \label{app:firstlocked}
When $\beta=0$, it is possible to calculate the chimera state
exactly, at the moment of its birth from a spatially modulated
drift state. Recall that all states of pure drift satisfy
$\Theta(x) \equiv 0$, and equivalently, that $a$ has zero
imaginary part.  Hence we can seek solutions of the algebraic
self-consistency equations with real values of $a$ and $c$, for
$\beta=0$.  At the onset of the chimera, the first locked
oscillators are born.  As suggested by
Fig.~\ref{fig:newGdynamics}, this occurs when the graph of $R(x)$
intersects the horizontal line $R=\Delta$ tangentially.

Therefore the bifurcation condition is $\Delta = R_{max}=c+a$.
Plugging this into \eqref{eq:c} and \eqref{eq:a}, and using
$\Theta(x) \equiv 0$ and $\beta=0$, we find that the
self-consistency equations become
\begin{equation} \label{eq:bifc2}
    c = \avg{\frac{c+a - \sqrt{(c+a)^2 - (c+a\cos x)^2}}{c+a\cos x}}
\end{equation}
and
\begin{equation} \label{eq:bifa2}
    a = A\avg{\frac{c+a - \sqrt{(c+a)^2 - (c+a\cos x)^2}}{c+a\cos x}\cos x}~.
\end{equation}
Note that both of these expressions can be rewritten solely in
terms of the ratio $a/c$, which suggests a neat way to solve them
parametrically.

Set $s=a/c$ and substitute into \eqref{eq:bifc2} above, which
becomes \begin{eqnarray} \label{eq:cofs}
    c &=& \avg{\frac{1+s - \sqrt{(1+s)^2 - (1+s\cos x)^2}}{1+s\cos x}} \nonumber \\
      &=& f_1(s)~.
\end{eqnarray}
So we can also write $a = sc = sf_1(s)$.

Similarly, the $a$ equation \eqref{eq:a} becomes:
\begin{eqnarray}
\label{eq:aofs}\
    a &=& A\avg{h\cos x} \nonumber \\
      &=& A\avg{\frac{1+s - \sqrt{(1+s)^2 - (1+s\cos x)^2}}{1+s\cos x}\cos x} \nonumber \\
      &=& A f_2(s)~.
\end{eqnarray}

All other quantities of interest can also be expressed in terms of
$s$.  For instance, we can now substitute $a = sc = sf_1(s)$ into
\eqref{eq:aofs} and solve for $A(s) = a/f_2(s) = s f_1(s)/f_2(s)$.
Likewise, $\Delta=c+a = (1+s)f_1(s)$. In summary, the incipient
chimera state can be written exactly in parametric form, as
follows: \begin{eqnarray} \label{eq:sfuncs}\
    c &=& f_1(s) \nonumber \\
    a &=& sf_1(s) \nonumber \\
    A &=& s\frac{f_1(s)}{f_2(s)} \nonumber \\
    \Delta &=& (1+s)f_1(s)~.
\end{eqnarray}

Since $A$ is a control parameter of the original equations (the
only free one after $\beta$ has been chosen to be zero), it is
desirable to reparametrize this solution in term of $A$.  To do
that, we invert $A(s)$ in \eqref{eq:sfuncs} to obtain the
following series expansion for $s(A)$,
\[
s \sim \frac{16}{9\pi^2} A^2 - \frac{16}{27}
\paren{\frac{3\pi^2-32}{27\pi^4}}A^3 + \order{A^4}~,
\]
and use that to rewrite the newborn chimera in terms of
$A$:
\begin{eqnarray*}
    c &\sim& 1 - \frac{16}{3\pi^2}A + \frac{8}{9} \paren{\frac{5\pi^2-32}{\pi^4}}A^2 +
        \order{A^3}~,\\
    a &\sim& \frac{16}{9\pi^2}A^2 - \frac{16}{27} \paren{\frac{3\pi^2-16}{\pi^4}}A^3 +
        \order{A^4}~,\\
    \Delta &\sim& 1 - \frac{16}{3\pi^2}A + \frac{8}{9} \paren{\frac{7\pi^2-32}{\pi^4}}A^2
        + \order{A^3}~.
\end{eqnarray*}
Notice that this has exactly the form of the ansatz we postulated
in \eqref{eq:ansatz}, based on numerical experiments. As expected,
it satisfies $\Delta_1=c_1$ as in \eqref{eq:loworder} and gives
$\Delta_2-c_2 = 16/(9 \pi^2)$.

\newpage

\bibliographystyle{plain} 
\bibliography{allrefsforpaper}

\end{document}